\documentclass[
 ,final            % use final for the camera ready runs
  ]
  {aipproc}

\layoutstyle{8x11double}

\usepackage{color}

\def\gsim{\:\raisebox{-0.5ex}{$\stackrel{\textstyle>}{\sim}$}\:}

\newcommand{\newc}{\newcommand}

\newcommand{\beq} {\begin{equation}}
\newcommand{\eeq} {\end{equation}}
\newcommand{\bea}{\begin{eqnarray}}
\newcommand{\eea}{\end{eqnarray}}

\newc{\ssup}{\tilde{u}}
\newc{\ssdown}{\tilde{d}}
\newc{\ssstrange}{\tilde{s}}
\newc{\sscharm}{\tilde{c}}
\newc{\sstop}{\tilde{t}}
\newc{\ssbottom}{\tilde{b}}
\newc{\sse}{\tilde{e}}
\newc{\ssmu}{\tilde{\mu}}
\newc{\sstau}{\tilde{\tau}}
\newc{\ssnue}{\tilde{\nu}_{e}}
\newc{\ssnumu}{{\tilde{\nu}_{\mu}}}
\newc{\ssnutau}{{\tilde{\nu}_{\tau}}}
\newc{\ssbnue}{\tilde{\nu}^*_{e}}
\newc{\ssbnumu}{\tilde{\nu}^*_{\mu}}
\newc{\ssbnutau}{\tilde{\nu}^*_{\tau}}
\newc{\neut}{{\tilde{\chi}}^0}
\newc{\charge}{\tilde{{\chi}}}
\newc{\glu}{\tilde{g}}
\newc{\Higgs}{H^0}
\newc{\Azero}{A_0}

\newc{\muon}{\mu}
\newc{\azero}{A_0}
\newc{\neutralino}{\tilde\chi^0}
\newc{\selectron}{\tilde e}
\newc{\stau}{\tilde\tau}
\newc{\smuon}{\tilde\mu}
\newc{\sneu}{\tilde\nu}
\newc{\higgs}{h^0}
\newc{\sgnmu}{\textrm{sgn}(\mu)}
\newc{\gev}{\mbox{~GeV}}
\newc{\tev}{\mbox{~TeV}}
\newc{\mgut}{{M_{\rm GUT}}}
\newc{\mweak}{{M_{W}}}

\newc{\lam}{\lambda}
\newc{\lamp}{\lambda^\prime}
\newc{\lampp}{\lambda^{\prime\prime}}
\newc{\Lam}{\Lambda}
\newc{\BLam}{{\mathbf{\Lam}}}
\newc{\eps}{\epsilon}
\newc{\kap}{\kappa}

\newc{\del}{\partial}
\newc{\veva}{\langle H_1\rangle}
\newc{\vevb}{\langle H_2\rangle}
\newc{\onehalf}{\textstyle \frac{1}{2} \displaystyle}
\newc{\onethird}{\textstyle \frac{1}{3} \displaystyle}
\newc{\mzero}{M_0}
\newc{\mhalf}{{M_{1/2}}}
\newc{\tanb}{\tan\beta}
\newc{\Psix}{{{P}_{\!6}}}
\newc{\nPsix}{{\not\!\Psix}}
\newc{\nPsixU}{{\s{P}_{\!6}}}

\begin{document}

\title{Sneutrino LSPs in R-parity violating minimal supergravity models}


%04.65.+e Supergravity
%12.60.Jv Supersymmetric models
%14.80.Ly Supersymmetric partners of known particles

\author{Markus A. Bernhardt}{
 address={ Physikalisches Institut der Universit\"at Bonn,
 Nu\ss allee 12, D-53115 Bonn, Germany } 
}

\author{Siba Prasad Das}{
 %  address={<common address for author2 and author3>}
}

\author{Herbi K. Dreiner}{
%  address={<common address for author1 and author3>}
%  ,altaddress={<author1 address>} % additional visiting address
}

\author{Sebastian Grab}{
%  address={<common address for author2 and author3>}
}

\begin{abstract}
We consider the minimal supergravity model (mSUGRA) with one
 additional R-parity violating operator at the GUT scale. The
 superparticles mass spectra at the weak scale are generally altered
 due to the presence of the R-parity violating coupling in the
 renormalization group equations. We show that a lepton number
 violating coupling at the GUT scale can lead to a sneutrino
 as the lightest supersymmetric particle (LSP) in a large region
 of parameter space consistent with the muon anomalous magnetic 
 moment and other precision measurements. We also give characteristic 
 collider signatures at the LHC.
\end{abstract}

\maketitle

%%%%%%%%%%%%%%%%%%%%%%%%%%%%%%%%%%%%%%%%%%%%
%% MAINMATTER
%%%%%%%%%%%%%%%%%%%%%%%%%%%%%%%%%%%%%%%%%%%%

\section{Introduction}
%%%%%%%\section{Explicit R-parity violation in mSUGRA}

The most general gauge invariant and renormalizable superpotential 
of the supersymmetric standard model with minimal particle content 
contains lepton and baryon number violating operators
\cite{Allanach:2003eb},      
\bea
W_{\not {R}_p} &=&
 \, \frac{1}{2}\lambda_{ijk} L_iL_j{\bar E}_k + \lambda_{ijk}^\prime L_iQ_j{\bar D}_k  
\nonumber \\
& & + \, \frac{1}{2}\lambda_{ijk}^{\prime\prime}{\bar U}_i{\bar D}_j{\bar D}_k + \kappa_i L_iH_u , 
\label{notP6-superpot}
\eea
where $i,j,k=1,2,3$ are the generation indices. 

The fourth term in Eq.~(\ref{notP6-superpot}) can be rotated away at any scale by  an 
appropriate field redefinitions~\cite{Hall:1983id}.   
Simultaneous presence of lepton (first two terms) and baryon (third term) number violating operators
lead to rapid proton decay \cite{Barbier:2004ez}.
An additional discrete symmetry is therefore required to keep the proton stable
\cite{discrete_symmetries}. R-parity, $R_p$, and proton-hexality, $P_6$,
prohibits $W_{\not R_p}$. Here, we will consider a third possibility, 
baryon-triality, $B_3$, which violates $R_p$ and $P_6$ by prohibiting 
only the $\bar{U}\bar{D}\bar{D}$ operator in Eq.~(\ref{notP6-superpot}).

The remaining two lepton number violating interactions in Eq.~(\ref{notP6-superpot}) 
will lead to the decay of the lightest supersymmetric particle (LSP).  
Since the LSP is not stable, any SUSY particle (sparticle) is allowed to
be the LSP \cite{Ellis:1983ew}. However, the nature of the LSP is crucial for supersymmetric
signatures at colliders, since typically all heavier sparticle decay chains   
end up with the LSP. 
%Moreover, the final state topology depends upon the  
%sparticles mass ordering. It is difficult to study in detail  
%all possible mass orderings of the sparticles. We thus need well
%motivated models for detailed phenomenological studies.

\smallskip

In the $B_3$ minimal supergravity (mSUGRA) model \cite{Allanach:2003eb}, 
we assume an additional coupling $\Lambda'$ at the GUT scale ($M_{\rm GUT}$). 
We thus have the six free parameters:
\bea
&&\mzero\,,\, \mhalf\,,\,
\azero\,,\,\tanb\,,\, \sgnmu\,  {\rm ~and~}  \,\Lambda' \quad,
\label{P6V-param}
\eea
where $\Lambda'$ is one coupling out of $\lambda'_{ijk}$. Due to the presence 
of one such coupling, the collider phenomenology could be very different  
compared to $R_p$ conserving mSUGRA scenarios with a stable neutralino LSP.  
The possible changes are the following: different running of sparticle masses and  
couplings \cite{Allanach:2003eb,Allanach:2006st,Jack:2005id}; 
the LSP will decay; single production of sparticles \cite{Barbier:2004ez,Dreiner:2008rv}; 
the decay  patterns of the sparticles are altered \cite{Barbier:2004ez,Allanach:2006st,Dreiner:2008rv}.

It was shown in Ref.~\cite{Allanach:2006st} that the model  
possess three different LSP candidates: the lightest
neutralino, $\neut_1$, the lightest scalar tau (stau),
$\tilde{\tau}_1$, and the sneutrino, $\tilde{\nu}_i$. The $\tilde{\nu}_i$ is special,
because its mass needs to be reduced by additional $B_3$
contributions in the renormalization group equations (RGEs) in order
to become the LSP. We will mainly focus on that and show that 
a large region in the
$B_3$ mSUGRA parameter space exists, where a $\tilde{\nu}_i$ is the LSP 
consistent with experimental constraints, $e.g.$, 
the anomalous magnetic moment of the muon, $a_\mu$ \cite{Stockinger:2007pe} and  
the LEP bounds on lighter Higgs boson mass and sneutrino masses \cite{LEP_bounds}.

\section{Sneutrino LSPs in $B_3$ mSUGRA}
\label{snu_LSP_in_mSUGRA}

The dominant contributions to the RGE of $m_{\tilde{\nu}_i}$
can be expressed as \cite{Allanach:2003eb}: 
\bea
16\pi^2 \frac{d(m^2_{\tilde{\nu}_i})}{dt} &=&
- \, \Big( \frac{6}{5} g_1^2 |M_1|^2 + 6 g_2^2 |M_2|^2
+ \frac{3}{5} g_1^2  S \Big) \nonumber \\
& & + \, 6\lambda'^2_{ijk}\left[ ({\bf m_{\tilde{L}}})^2_{ii}
+({\bf m_{\tilde{Q}}})^2_{jj} +({\bf m_{\tilde{D}}})^2_{kk} \right]
\nonumber \\
& & + \, 6 ({\bf h_{D^k}})_{ij}^2
\label{sneu_RGE}
\eea
with
\bea
({\bf h_{D^k}})_{ij} \equiv \lambda'_{ijk} \times A_0 \,\,\, {\rm ~at~} \,\,\, M_{\rm GUT}.
\label{hdk_RGE}
\eea

The running of $m_{\tilde{\nu}_i}$ is governed by two different terms. The first 
term in Eq.~(\ref{sneu_RGE}) is  proportional to the gauge couplings $g_1$, $g_2$ 
and gaugino mass parameters $M_1$, $M_2$. $S$ is identically zero at $M_{\rm GUT}$   
due to the universal nature of scalar masses in mSUGRA,  
see Ref.~\cite{Allanach:2003eb}. For vanishing $\lambda'_{ijk}$, 
$m_{\tilde{\nu}_i}$ increases while running from $M_{\rm GUT}$ to the electroweak 
scale ($M_Z$) due to the presence of the negative sign in front of the gauge terms.
     
The contribution from the second term is proportional 
to $\lambda'_{ijk}$ and $({\bf h}_{D^k})_{ij}$, which is also proportional
to $\lambda'_{ijk}$ at $M_{\rm GUT}$, {\it cf.} Eq.~(\ref{hdk_RGE}). 
This terms  are, in contrast to the gauge terms, positive and will therefore   
reduce $m_{\tilde{\nu}_i}$ when we go from $M_{\rm GUT}$ to $M_Z$. 
The influence of these new contributions on $m_{\tilde{\nu}_i}$ 
depend on the magnitude of $\lambda'_{ijk}$ and also on the other mSUGRA parameters, 
especially on $A_0$ \cite{ddbg}.

\begin{figure}
  \includegraphics[height=.27\textheight,width=0.48\textwidth]{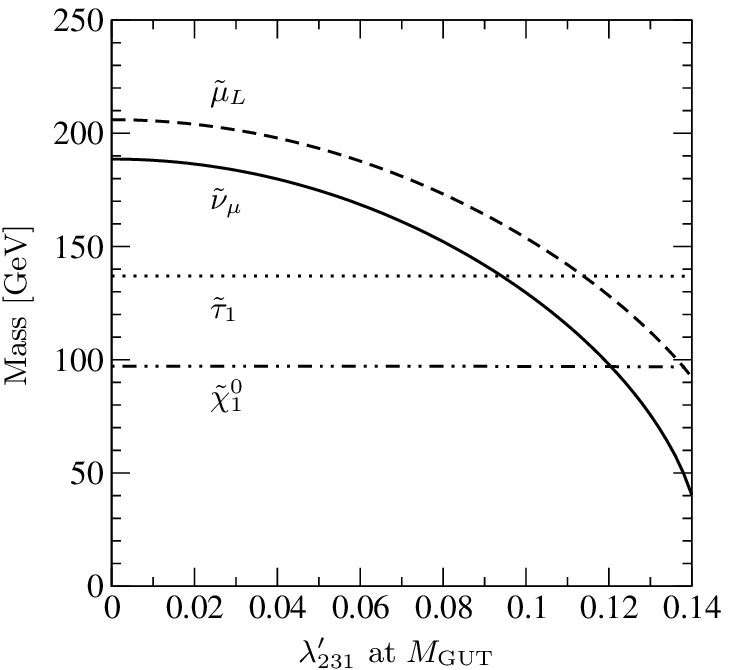}
%\put(-231.2,59.5){\rotatebox{90}{{ Mass [GeV]}}}
%\put(-120.0,0.3){$\lambda'_{231}$ at $M_{\rm GUT}$}
%\put(-180.0,63.5){$\neutralino_1$}
%\put(-180.0,87.5){$\sstau_1$}
%\put(-180.0,114.5){$\tilde{\nu}_\mu$}
%\put(-180.0,128.5){$\tilde{\mu}_L$}
%\put(-5.2,4.3){$\tilde{\nu}_\mu$}
%  \put(-5.2,5.4){$\tilde{\mu}_L$}
%  \put(-5.2,3.2){$\sstau_1$}
%  \put(-5.2,2.3){$\neutralino_1$}
  \caption{{\small  
Masses of $\neutralino_1$, $\sstau_1$, $\ssnumu$ and $\tilde{\muon}_L$ at $M_Z$
as a function of $\lambda'_{231}|_{\rm GUT}$ for the SPS1a mSUGRA parameters.}}
\label{fig:sps1amasses_Lamp231}
\end{figure}
\begin{figure}
  \includegraphics[height=.27\textheight,width=0.48\textwidth]{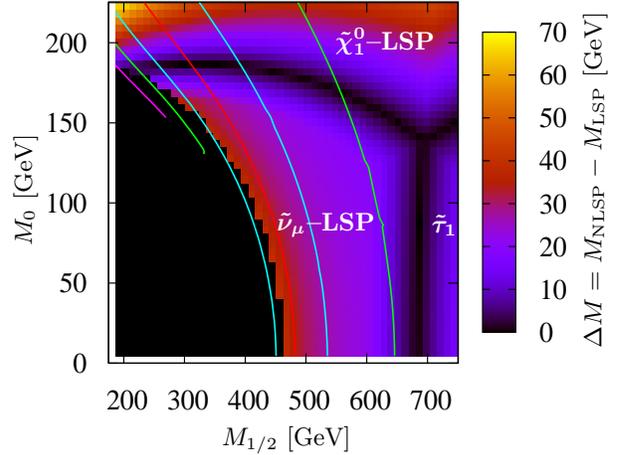} 
%\put(-5.2,39.5  ){\rotatebox{90}{$\Delta M = M_{NLSP} - M_{LSP}$ [GeV]}}
%\put(-120.0,0.3){\makebox(0,0){$\mhalf$ [GeV]}}
%\put(-231.2,69.5){\rotatebox{90}{$\mzero$ [GeV] }}
  \caption{{\small  
Mass difference, $\Delta M$, between the LSP and NLSP.  
The LSP candidates in different regions are explicitly mentioned.
The blackened out region corresponds to parameter points, which possess a 
tachyon or where the $\tilde{\nu}_\mu$ or the Higgs boson mass violates the LEP bounds   
of Ref.~\cite{LEP_bounds}.}}
\label{fig:m12m0_delta_Lamp231}
\end{figure}

We show in Fig.~\ref{fig:sps1amasses_Lamp231} the impact of a 
non-vanishing coupling $\lambda'_{ijk}|_{\rm GUT}$ on the running of $m_{\tilde{\nu}_i}$  
for the SPS1a parameter set \cite{Allanach:2002nj}. For this purpose we use
an unpublished $B_3$ version of {\tt SOFTSUSY} \cite{rpv_softsusy}. 
We observe, that $m_{\tilde{\nu}_{\mu}}$ is decreasing for increasing 
$\lambda'_{231}|_{\rm GUT}$ as described by Eq.~(\ref{sneu_RGE}).
Furthermore, the mass of the left-handed smuon, $\tilde{\mu}_L$, also decreases because 
it belongs to the same SU(2) doublet like the $\tilde{\nu}_\mu$. 
The masses of the $\tilde{\chi}_1^0$  
and $\tilde{\tau}_1$ are almost insensitive to $\lambda'_{231}|_{\rm GUT}$. 
We therefore obtain a $\tilde{\nu}_\mu$-LSP for $\lambda'_{231} > 0.12|_{\rm GUT}$.
This large coupling is still consistent with experimental observations
\cite{Barbier:2004ez}. 

In general, there are large regions in the $B_3$ mSUGRA parameter space
where the $\tilde{\nu}_i$ becomes the LSP, if $\lambda'_{ijk}|_{\rm GUT} \gsim {\cal O}(0.1)$.
As an example we choose $A_0$ = -600 GeV, $\tan\beta$=10, {\rm sgn}$(\mu)$ = +1, $\lamp_{231} =0.11|_{\rm GUT}$  
and vary $M_0$ and $M_{1/2}$. We show in Fig.~\ref{fig:m12m0_delta_Lamp231} 
the mass difference between the LSP and the next-to-LSP (NLSP) in the 
$M_0$--$M_{1/2}$ plane.  We observe three different LSP candidates, namely the 
$\tilde{\nu}_\mu$, the $\neutralino_1$ and the $\sstau_1$, in different 
regions of $B_3$ mSUGRA parameter space. Increasing $M_{1/2}$ increases the
mass of the (left-handed) $\tilde{\nu}_\mu$ faster than the mass of the
(mainly right-handed) $\sstau_1$. We thus find a $\sstau_1$-LSP instead of a 
$\tilde{\nu}_\mu$-LSP for large $M_{1/2}$. Increasing $M_0$, increases the 
masses of the sfermions but not the mass of the (bino-like) $\neutralino_1$
and we therefore reobtain the $\neutralino_1$-LSP.  
We also see that the LSP and NLSP are almost degenerate in mass around the thick black 
contours. 

We conclude that a $\tilde{\nu}_\mu$-LSP exists in an extended region of $B_3$ 
mSUGRA parameter space, which is also consistent with the observed anomalous magnetic moment
of the muon, $a_\mu$. The contours in Fig.~\ref{fig:m12m0_delta_Lamp231} correspond   
to the measured central value of $a_\mu$ (red), $\pm 1 \sigma$ (green), 
$ \pm 2 \sigma$ (blue) and $\pm 3 \sigma$ deviation (magenta)
from the central value. The SUSY contributions to $a_\mu$ are
calculated using {\tt micrOMEGAs1.3.6} \cite{Belanger:2001fz}.

%%%%%%%%%%%%%%%%%%%%%%%%%%%%%%%%%%%%%%%%%%%%
%% SAMPLE TABLE
%%
%% Shows the use of \tablehead and \tablenote
%% macros
%%%%%%%%%%%%%%%%%%%%%%%%%%%%%%%%%%%%%%%%%%%%

\section{LHC Phenomenology}

We choose $M_0$=100 GeV and $M_{1/2}$=450 GeV from Fig.~\ref{fig:m12m0_delta_Lamp231} as a  
representative point for investigating the $\tilde{\nu}_\mu$-LSP    
phenomenology at the LHC. We calculate the decay rates with {\tt ISAWIG1.200} and 
{\tt ISAJET7.64} \cite{Paige:2003mg}. This output is fed into {\tt HERWIG} 
\cite{Herwig} to simulate events at the LHC. We show some masses and BRs,
which are special for $\tilde{\nu}_\mu$-LSP scenarios in  
Table.~\ref{tab:massBR}. 

\begin{table}
%%%%%%%%%%%%%%%%\scalebox{0.69}{
\begin{tabular}{lrrrrrr}
\hline
  & \tablehead{1}{r}{b}{ mass}
  & \tablehead{1}{r}{b}{channel}
  & \tablehead{1}{r}{b}{BR} 
  & \tablehead{1}{r}{b}{channel}
  & \tablehead{1}{r}{b}{BR}\\
\hline
$\ssnumu$ & {\bf 124} & $\bar b d$  & {\bf 100\% }&&\\
\hline
$\ssmu^-_L$ & {\bf 147} & $W^- \bar b d$ & {\bf 75.1\%} &  $\bar c d$ & {\bf 24.9\%} \\
\hline
$\neut_1$ & 184 & $\ssnumu^* \nu_\mu$   &  36.0\% & $\ssnumu \bar \nu_\mu$ & 36.0\%  \\
& & $\ssmu^+_L \mu^-$ &14.0\%  & $\ssmu^-_L \mu^+$ &  14.0\%  \\
\hline
$\sstop_1$ & {\bf 650} & $\charge^+_1 b$   & $42.1\%$ &  $\neut_1 t$ & $33.5\%$ \\
& & $\neut_2 t$ & $13.8\%$ & $\mu^+ d$ & {\bf 10.6}$\%$  \\
\hline
$\tilde d_R$ & {\bf 897} &$\nu_\mu b$ & {\bf 45.3}$\%$
& $\mu^- t$ & {\bf 42.1}$\%$ \\
& &$\neut_1 d$ & $12.6\%$ & & \\
\hline
\end{tabular}
%%%%%%%%%%%%%%%%%%%%%%%%} % scalebox
\caption{  
Sparticle masses and branching ratios (BRs) in  
$B_3$ mSUGRA models. Masses which are affected  
due to $\lambda'_{231}$ coupling and $B_3$ decays are shown in bold-face.}
\label{tab:massBR}
\end{table}

\begin{figure}
  \includegraphics[height=.3\textheight,width=0.48\textwidth]{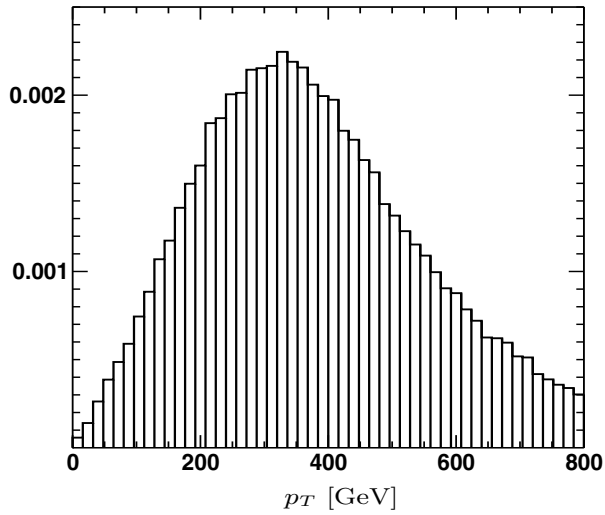}
%\put(-120.0,0.3){\makebox(0,0){$p_T$ [GeV]}}
%\put(-241.2,69.5){\rotatebox{90}{ $ \frac{1}{\sigma} \frac{d\sigma}{d{p_T}}$ $\times GeV$}}
%    \put(-107.0,-13.0){$p_T$ [GeV]}
%  \put(-207.0,81.0){\large{\rotatebox{90}{ $ \frac{1}{\sigma} \frac{d\sigma}{d{p_T}}$ $GeV^{-1}$ }}}
  \caption{{\small  
        The normalized $p_T$ distribution of the muon from the decays
        $\tilde{d}_R \rightarrow \mu t $ and
        $\tilde{t}_{1/2} \rightarrow \mu d$ ({\it cf}. Table.~ 
        \ref{tab:massBR}) at the LHC.
}}
\label{fig:ptMuonfromdRtLR_Lamp231}
\end{figure}
The sparticle pair production cross-section   
at the LHC is 3.0 pb. Once produced, the sparticles might 
follow the decays in Table.~\ref{tab:massBR}. This will lead to 
very distinctive final states compared to mSUGRA with a stable 
$\tilde{\chi}_1^0$-LSP.  

One of the most striking signatures of this scenario are high-$p_T$ muons 
($\approx$  10\% of events) from the direct decays of $\tilde{d}_R$ and $\tilde{t}_{1/2}$ via 
$\lambda'_{231}$, see Table.~\ref{tab:massBR}. We show the normalized distribution in   
Fig.~\ref{fig:ptMuonfromdRtLR_Lamp231}. The distribution peaks at 300 GeV and   
one can thus apply a very stringent selection cut on the muon $p_T$ 
to isolate the SUSY signal from the SM background. 

Additional distinctive signatures are ({\it cf.} Table.~\ref{tab:massBR}): 
not necessarily missing $p_T$ ($\approx$  25\% of events), because the $\tilde{\nu}_\mu$-LSP  
and the $\tilde{\mu}_L$-NLSP will decay into SM particles; 2-5 non-$b$ jets and 0-4 $b$ jets, where most 
of the $b$ jets stem from the LSP decay; high-$p_T$ top quarks from 
$\tilde{d}_R$ decay; like-sign dimuon events mainly from pair production 
of right-handed squarks and subsequent decays into the (Majorana) 
$\tilde{\chi}_1^0$ and $\tilde{\mu}_L$. We will present the detailed 
analysis in \cite{ddbg}.    

% Another unique possibility of $B_3$ model is the single sparticle production, $e.g.$, $\tilde{\nu}_\mu + X$ 
% followed by the decay in $\bar b d$ channel ({\it cf.} Table.~\ref{tab:massBR}) leads to di-jet signature. The   
% cross-sections depends directly on the magnitude of 
% the $\not P_6$ coupling and is $\approx$ $2.2 \times 10^{6}$ fb at LHC for typical $\tilde{\nu}_\mu$-LSP 
% scenario. However, it suffers large QCD backgrounds.

\section{conclusions}

We have shown that a coupling $\lambda'_{231}|_{\rm GUT}=\mathcal{O}(10^{-1})$ 
can lead to a sneutrino LSP in $B_3$ mSUGRA models (Fig.~\ref{fig:sps1amasses_Lamp231}). 
We have found large regions of parameter space where the sneutrino is the LSP 
consistent with $a_\mu$ and other electroweak precision observables 
(Fig.~\ref{fig:m12m0_delta_Lamp231}). Distinctive LHC signatures 
are high-$p_T$ muons (Fig.~\ref{fig:ptMuonfromdRtLR_Lamp231}), 
high-$p_T$ tops, like-sign dimuon events and $b$ jets.

%%%%%%%%%%%%%%%%%%%%%%%%%%%%%%%%%%%%%%%%%%%%%%%%
%% BACKMATTER
%%%%%%%%%%%%%%%%%%%%%%%%%%%%%%%%%%%%%%%%%%%%%%%%

\begin{theacknowledgments}
We thank Ben Allanach for help with the yet unpublished $B_3$ version of {\tt SOFTSUSY}.     
SPD acknowledges financial support from the Bundesministerium f\"ur Bildung
und Forschung (BMBF) Projekt under Contract No. 05HT6PDA. SG thanks the 
"Deutsche Telekom Stiftung" and the "Bonn Cologne Graduate School" for financial support.

\end{theacknowledgments}

\end{document}